\documentclass[%
  manuscript=letter,
  journal=jpclcd
]{achemso}

\usepackage[utf8]{inputenc}
\usepackage[T1]{fontenc}
\usepackage{xcolor}
\definecolor{blue}{RGB}{0, 0, 139}
\usepackage[%
  colorlinks=true,
  allcolors=blue
]{hyperref}
\usepackage{url}
\usepackage{enumitem}
\usepackage{amsfonts}
\usepackage{amssymb}
\usepackage{amsmath}
\usepackage{mathtools}
\usepackage{lmodern}

\DeclarePairedDelimiter\ppar{(}{)}              % ( )
\DeclarePairedDelimiter\pnrm{\lVert}{\rVert}    % || ||
\DeclarePairedDelimiter\pbkt{[}{]}              % [ ]
\DeclarePairedDelimiter\pset{\{}{\}}            % { }

\newcommand{\rfig}[1]{Figure~\ref{#1}}
\newcommand{\rref}[1]{ref~\citenum{#1}}
\newcommand{\req}[1]{eq~\ref{#1}}

\newcommand{\bx}{\mathbf{x}}

\newcommand{\e}{\operatorname{e}}

\newcommand{\note}[1]{{\color{black}{#1}}}

\title{\Large Selecting High-Dimensional Representations of Physical Systems by Reweighted Diffusion Maps}

\author{Jakub Rydzewski}
\email{jr@fizyka.umk.pl}
\affiliation{%
  Institute of Physics,
  Faculty of Physics, Astronomy and Informatics,
  Nicolaus Copernicus University,
  Grudziadzka 5, 87-100 Toru\'n, Poland
}

\begin{document}

\begin{tocentry}
  \begin{center}
    \includegraphics{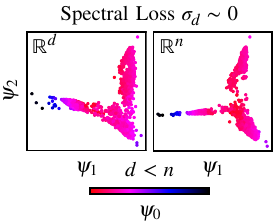}
  \end{center}
\end{tocentry}

\begin{abstract}
Constructing reduced representations of high-dimensional systems is a fundamental problem in physical chemistry. Many unsupervised machine learning methods can automatically find such low-dimensional representations. However, an often overlooked problem is what high-dimensional representation should be used to describe systems before dimensionality reduction. \note{Here, we address this issue using a recently developed method called reweighted diffusion map [{\it J. Chem. Theory Comput.} {\bf 2022}, {\it 18}, 7179--7192].} We show how high-dimensional representations can be quantitatively selected by exploring the spectral decomposition of Markov transition matrices built from data obtained from standard or enhanced sampling atomistic simulations. We demonstrate the performance of the method in several high-dimensional examples.
\end{abstract}

\maketitle

Machine learning is becoming widely used for performing dimensionality reduction in atomistic simulations and general analysis of high-dimensional systems in physical chemistry. Many unsupervised methods have been developed to extract a few variables encoding macroscopic information about complex processes hidden in high-dimensional observations from simulation data~\cite{nadler2006diffusion,tribello2012using,hashemian2013modeling,rohrdanz2013discovering,rydzewski2016machine,mccarty2017variational,rydzewski2018kinetics,zhang2018unfolding,rodriguez2019identifying,rydzewski2020multiscale,morishita2021time,rydzewski2022reweighted}. \note{The quality of variables resulting from such methods depends heavily on the input data consisting of many configuration variables referred to as features.} However, the selection of such high-dimensional representations used subsequently for dimensionality reduction is often overlooked or performed by trial and error.

A possible basis for the interpretable construction of high-dimensional representations in complex systems is the preservation of their timescale separation between slow and fast variables. The slow variables are intrinsically related to the kinetics of rare transitions between long-lived metastable states in configuration space~\cite{valsson2016enhancing,bussi2020using}, which are essential in many processes, for instance, catalysis~\cite{piccini2022ab}, crystallization~\cite{neha2022collective}, or conformational transitions~\cite{rydzewski2017ligand,van2021towards}. The fast variables, however, are adiabatically slaved to the dynamics of the slow variables and correspond mainly to equilibration within metastable states. Therefore, we can consider different representations of the same system equivalent if the same timescale separation characterizes them.

In this Letter, we exploit this idea and develop a method for a quantitative and interpretable selection of the high-dimensional configuration space based on the spectral decomposition of Markov transition matrices from simulation data. Eigendecomposition has a long history of being used to analyze complex high-dimensional spaces, particularly for finding a low-dimensional manifold on which the data resides~\cite{berard1994embedding,belkin2001laplacian,belkin2003laplacian,mezic2005spectral,coifman2005geometric,coifman2006diffusion,jones2008manifold,coifman2008diffusion,tiwary2016spectral}. In contrast to such approaches, our method does not use eigenfunctions to parametrize the slow coarse-grained variables. Instead, it iteratively removes variables from the complete high-dimensional representation to find a partial selection of configuration variables while preserving kinetic information about the complete high-dimensional representation.

Let us first introduce the concept of the high-dimensional representation of physical systems. Each microscopic configuration of the system is described by $n$ configuration variables (i.e., features) $\bx = \ppar*{x_1, \dots, x_n}$. In the case of the microscopic coordinates, the configuration variables are sampled from an equilibrium probability distribution given by the Boltzmann density $p(\bx) \propto \e^{-\beta U(\bx)}$, where $U(\bx)$ is the potential energy of the system, and $\beta=\ppar{k_{\mathrm{B}}T}^{-1}$ is the inverse temperature. However, the equilibrium probability distribution is usually unknown for other high-dimensional spaces (e.g., invariant representations).

To estimate the kinetic information encoded in the high-dimensional space, we collect $N$ samples of $n$ configuration variables from a simulation to construct the Markov transition matrix and perform its spectral decomposition. To this aim, a data set consisting of these samples is:
\begin{equation}
  \label{eq:data}
  X=\pset*{\bx_k \in \mathbb{R}^n, w(\bx_k)}_{k=1}^N,
\end{equation}
where the samples are augmented by statistical weights $w$ if we sample a biased probability distribution, such as in enhanced sampling simulations. The weights are given as follows:
\begin{equation}
  \label{eq:weight}
  w(\bx_k) \propto \frac{p(\bx_k)}{q(\bx_k)},
\end{equation}
where $p(\bx_k)$ and $q(\bx_k)$ are the unbiased and biased probability distributions at the $k$-th sample from $X$, respectively. For unbiased simulations, the weights are reduced to unity as they are sampled from the equilibrium distribution.

Next, let us introduce reweighted diffusion map~\cite{rydzewski2022reweighted}. We start by constructing an auxiliary Gaussian kernel which encodes information about the local geometry of the configuration space, $g_\varepsilon(\bx_k,\bx_l) = \exp\ppar{-\pnrm*{\bx_k-\bx_l}^2/2\varepsilon^2}$, where $\bx_k$ and $\bx_l$ are $n$-dimensional samples from the data set (\req{eq:data}) \note{and $\varepsilon$ is a scale constant which is chosen depending on the data set usually selected so that it matches the distance between neighboring samples. Here, we calculate the scale constant as the median of Euclidean distances $\pnrm{\cdot}$ for simplicity; see the Supplementary Information for details. Other techniques can also be used when further adjustment of the scale constant is required~\cite{coifman2008graph,berry2016variable,lindenbaum2020gaussian}}. 

\note{Then, a reweighted anisotropic kernel is introduced to employ additional information about the density and importance of the configuration space:
\begin{equation}
  \label{eq:diffusion-kernel}
  \kappa(\bx_k,\bx_l) = r(\bx_k,\bx_l)
  \frac{g_\varepsilon(\bx_k,\bx_l)}{\sqrt{\varrho(\bx_k) \varrho(\bx_l)}},
\end{equation}
where $\varrho(\bx)=\sum_{k}g_\varepsilon(\bx,\bx_k)$ is up to a multiplicative constant a kernel density estimate. In \req{eq:diffusion-kernel}, we reweight the anisotropic kernel using $r(\bx_k,\bx_l)$, which is introduced to correct the effect of sampling from the biased probability distribution $q$ (i.e., diffusion reweighting~\cite{rydzewski2022reweighted}). The reweighting factor is given as~\cite{rydzewski2020multiscale,rydzewski2022reweighted}:
\begin{equation}
  \label{eq:reweighting-factor}
  r(\bx_k,\bx_l) = \sqrt{w(\bx_k) w(\bx_l)},
\end{equation}
where $w(\bx_k)$ and $w(\bx_l)$ are the statistical weights corresponding to the $k$-th and $l$-th samples, respectively. See \rref{rydzewski2022reweighted} for a derivation of diffusion reweighting and a more general discussion. For unbiased simulations, \req{eq:reweighting-factor} reduces to the anisotropic diffusion kernel used in diffusion map~\cite{coifman2005geometric,singer2009detecting}.

Equation~\ref{eq:diffusion-kernel} asymptotically corresponds to a reversible, overdamped approximation to the slow dynamics with the unbiased probability $p(\bx)$ as the stationary density~\cite{coifman2005geometric,singer2009detecting} even if the underlying dynamics proceeds according to the biased probability distribution~\cite{rydzewski2022reweighted}. As such, the reweighted diffusion kernel given in \req{eq:diffusion-kernel} is a reasonable approximation for our method.}

Having calculated the reweighted anisotropic diffusion kernel (\req{eq:diffusion-kernel}), we can finally compute the related row-normalized kernel to define the Markov transition matrix $M(\bx_k,\bx_l)$ with the corresponding transition probabilities $m_{kl}$:
\begin{equation}
  \label{eq:markov-matrix}
  m_{kl} \sim M(\bx_k,\bx_l) = \frac{\kappa(\bx_k,\bx_l)}{\sum_n \kappa(\bx_k,\bx_n)}
\end{equation}
which is equivalent to an unbiased Markov chain given by $\operatorname{Pr}\pset*{\bx_{t+1}=\bx_l \; | \; \bx_t=\bx_k}$ that defines a probability that the system transitions from $\bx_k$ to $\bx_l$ in one timestep $t$. \note{Note that the time in the Markov chain is auxiliary and should not be confused with the simulation timestep or a timelag. Regardless of whether the data set is sampled from the biased probability distribution, the Markov transition matrix corresponds to the unbiased Markov chain after the reweighting~\cite{rydzewski2022reweighted}.}

\note{Reweighted diffusion map is very related to target measure diffusion map introduced by Banisch et al.~\cite{banisch2020diffusion}. The main difference netween both methods lies in their formulation. Here, we employ general statistical weights from an enhanced sampling simulation (\req{eq:weight}), and target measure diffusion map uses target probability distributions. Both techniques can be used in our framework. However, working with statistical weights is more suitable for our purposes as, usually, even approximate target measures are unknown.}

\note{Here, we use the Markov transition matrix $M$ to determine if the configuration space spanned by a partial selection of the configuration variables contains similar kinetic information as the high-dimensional representation of the system spanned by all its configuration variables.} For this, we first perform an eigendecomposition of the Markov transition matrix constructed from the complete set of the configuration variables $M\psi=\lambda\psi$ and calculate its eigenvalues $\pset{\lambda_k}$ and eigenfunctions $\pset{\psi_k}$. The eigenvalues are sorted by decreasing values. The corresponding eigenfunctions contain kinetic information about the system as the eigenvalues are related to the intrinsical timescales of the system.

Next, the eigendecomposition is carried out for combinations of the configuration variables, which define a data set $X_d$ ($d$ is the number of the configuration variables in the partial representation) and compare it to the eigenvalues of the complete high-dimensional representation. To describe how much kinetic information is conserved in the partial representations, we define a spectral loss:
\begin{equation}
  \label{eq:spectral-loss}
  \sigma_d = \alpha \pbkt*{\sum_k \ppar*{\lambda_{dk} - \lambda_k}^2}^{1/2},
\end{equation}
\note{where $\alpha$ is a normalization constant so that the value of the spectral loss for one variable is equal to 1}, and $\lambda_k$ and $\lambda_{dk}$ are the eigenvalues of the complete high-dimensional representation and the partial representation consisting of $d$ configuration variables, respectively. Therefore, a combination of the configuration variables preserves kinetic information encoded in the complete representation if the spectral loss is close to zero.

Given the data set $X$ of $n$ configuration variables $\bx = \ppar{x_1, \dots, x_n}$ and its spectral decomposition $\pset{\lambda_k, \psi_k}$ of the related Markov transition matrix $M$, we search for the partial high-dimensional data set $X_d$ of $d$ configuration variables that upon spectral decomposition of its Markov transition matrix into $\pset{\lambda_{dk}, \psi_{dk}}$ contains similar kinetic information as the Markov transition matrix calculated from $X$. To avoid an exhaustive and computationally demanding search through all combinations of the configuration variables, we use an algorithm that provides a suboptimal result~\cite{pudil1994floating}. Our algorithm is summarized below:
\note{
\begin{enumerate}
  \item[(1)] Start from the complete high-dimensional representation.
  \item[(2)] Repeat until the number of selected configuration variables is $d$:
  \begin{enumerate}
    \item[(a)] Remove from the data a variable corresponding to the minimal spectral loss upon removing one variable.
    \item[(b)] Add a configuration variable to the data if the spectral loss upon addition back decreases.
    \item[(c)] Go to (a) if adding any configuration variable does not result in decreasing the spectral loss; else, go to (b).   
  \end{enumerate}
  \item[(3)] Return the selected $d$-dimensional representation.
\end{enumerate}
}

\begin{figure}[t]
  \includegraphics{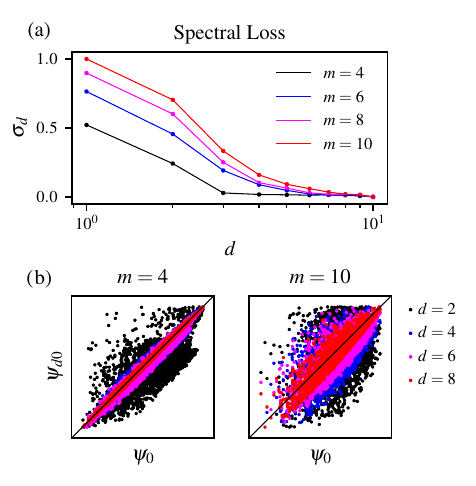}
  \caption{%
  High-dimensional data sets ($n=10$) sampled from multivariate Gaussian distributions with a different number of informative variables ($m$). For additional information about the benchmark data sets, see Supplementary Information.
  (a) Spectral loss $\sigma_d$ shown for the data sets consisting of $m=4, 6, 8, 10$ informative variables as a function of the number of the variables selected as the partial representation of the data set $d$. \note{Spectral loss is rescaled so that its value for one variable and $m=10$ is equal to one.}
  (b) Similarity between the equilibrium distributions calculated as the zeroth eigenfunctions for two selected data sets ($m=4$ and $m=10$). The eigenfunctions $\psi_{d0}$ are compared to $\psi_0$, i.e., the eigenfunctions of the Markov transition matrix $M$ computed from the complete data set $X$.
  }
  \label{fig:blob}
\end{figure}

As an initial test for our method, we apply it to four high-dimensional benchmark data sets of dimension $n=10$ sampled from multivariate Gaussian distributions that create clusters of samples on vertices of an $m$-informative hypercube with interdependence between these variables and additional noise. They consist of a different number of informative variables ($m=4, 6, 8, 10$). The remaining variables in these data sets are combinations of the informative variables. Further details about these data sets are available in Supplementary Information.

We present our results in \rfig{fig:blob}. By computing the spectral loss in reference to the eigendecomposition of the Markov transition matrix from the data set $X$, we can see that when the number of informative variables ($m$) is lower in the partial data sets, the data resembles $X$ much faster in comparison to the complete data set for $m=10$; see \rfig{fig:blob}(a). This observation indicates that the method correctly identifies the number of informative variables, as the improvement of the spectral representation of the high-dimensional space upon adding the remaining variables is negligible.

\note{This observation is also supported by checking how the eigenfunctions $\psi_{d0}$ of the partial high-dimensional representations converge to the eigenvalues calculated based on the Markov transition matrix from the data set $X$; see \rfig{fig:blob}(b)}. For example, the eigenfunctions $\psi_{n0}$ for the $m=4$ informative variables resemble $\psi_0$ much faster. In contrast, for $m=10$ informative variables, all the variables must be included in the data set $X_d$ to match the eigenvalues and eigenfunctions calculated from the Markov transition matrix from $X$.

\begin{figure*}[t]
  \includegraphics{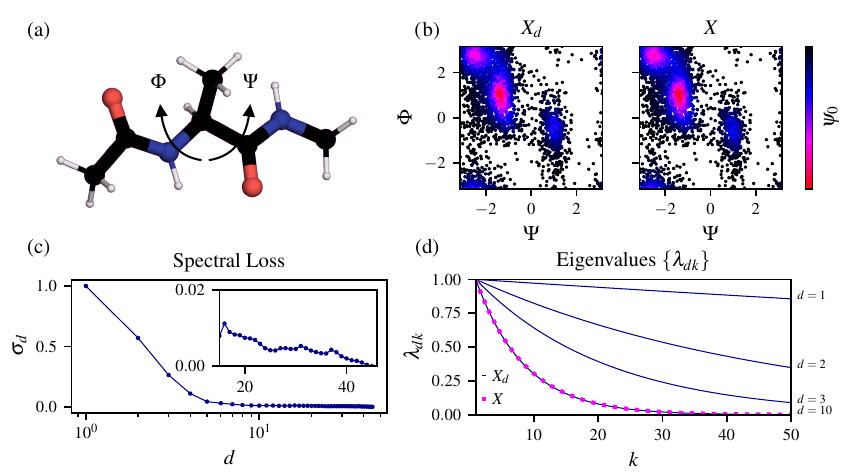}
  \caption{
  Selecting high-dimensional configuration space for alanine dipeptide (Ace-Ala-Nme) in vacuum.
  (a) Structure of alanine dipeptide with the $\Phi$ and $\Psi$ dihedral angles marked. The system is represented by distances between heavy atoms ($n=45$) from the data set of $N=2500$ samples generated using well-tempered metadynamics biasing $\Phi$ and $\Psi$ with a bias factor of 5. For additional information about the simulation, see Supplementary Information.
  (b) Equilibrium density calculated as the zeroth eigenvector of the Markov transition matrices using reweighted diffusion map in the space spanned by the $\Phi$ and $\Psi$ dihedral angles.
  (c) Spectral loss $\sigma_d$ showing the convergence of the partial selection of the configuration variables to the reference data set at about $d\sim{10}$ (the number of the configuration variables in the partial representations $d$ is shown on a logarithmic scale). \note{Spectral loss is rescaled so that its value for one variable is equal to one.}
  (d) Eigenvalues $\pset{\lambda_{dk}}$ ($k$ is the index of the eigenvalue) calculated from the partial data sets $X_d$ ($d=1,2,3,10$) shown in blue and $X$ shown in magenta. The last shown spectrum for $d=10$ is the same as for the complete data set $X$.
  }
  \label{fig:ala1}
\end{figure*}

As a subsequent example, we consider alanine dipeptide in vacuum, which has two main long-lived metastable states that its $\Phi$ dihedral angle can separate; see \rfig{fig:ala1}(a). As a high-dimensional representation, we select all heavy-atom distances (the number of variables $n=45$) monitored during a 100-ns long simulation. For the data set, we use the last 10 ns of the simulation, sampled every 4 ps. The total number of samples used to construct the Markov transition matrix is $N=2500$. The simulation is performed using \textsc{gromacs} 2019.2~\cite{gromacs} patched with a development version of the \textsc{plumed} plugin~\cite{plumed,plumed-nest}. The biased data set $X$ is sampled using well-tempered metadynamics~\cite{barducci2008well} at 300 K with a bias factor of 5. The fluctuations of the $\Phi$ and $\Psi$ dihedral angles of alanine dipeptide are biased. The statistical weights are calculated using the Tiwary--Parrinello reweighting algorithm~\cite{tiwary_rewt} suitable for a time-dependent bias potential. Further details about simulation parameters are available in Supplementary Information.

We present the results in \rfig{fig:ala1}. \note{We can see in \rfig{fig:ala1}(b) that the equilibrium density spanned by the $\Psi$ and $\Psi$ dihedral angles given by the zeroth eigenfunction $\psi_0$ (the left eigenfunction of $M$ approximating stationary distribution of the Markov chain) is qualitatively identical for the partial representation consisting of $d=10$ variables and the complete high-dimensional representation ($n=45$).} These equilibrium densities correctly identify the long-lived metastable states of alanine dipeptide. The spectral loss can also confirm this conclusion quantitatively; see \rfig{fig:ala1}(c). We can see that after including $d=10$ variables, which give a steep decrease in the spectral loss, the spectral loss remains constant, and the error compared to the complete representation is roughly zero; see inset in \rfig{fig:ala1}(c). Adding the remaining configuration variables does not improve the spectral representation.

\begin{figure*}[t]
  \includegraphics{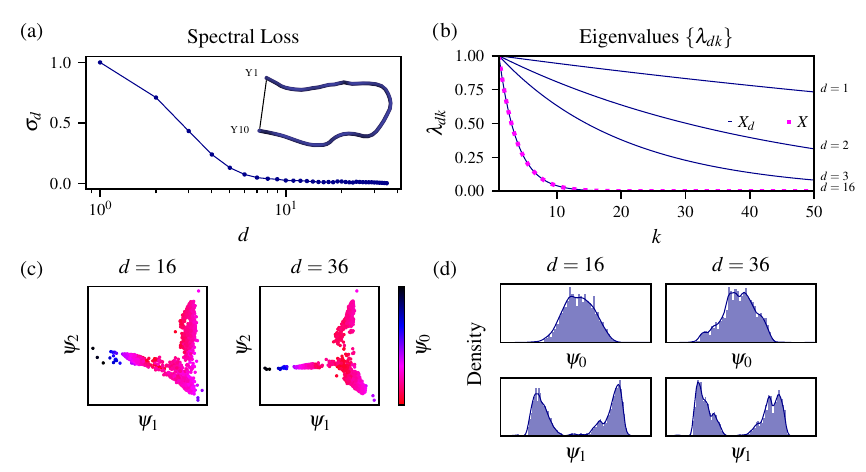}
  \caption{%
  Selecting high-dimensional configuration space for chignolin in solvent. The system is represented by the sines and cosines of the $\Phi$ and $\Psi$ dihedral angles ($n=36$). The data set consists of $N=2500$ samples generated using well-tempered metadynamics biasing the distance between the C$\alpha$ atoms of the Y1 and Y10 residues and the radius of gyration with a bias factor of 5 at 340 K. For details, we refer to Supplementary Information.
  (a) Spectral loss $\sigma_d$ for the partial high-dimensional representations of $X_d$ (the number of the configuration variables in the partial representations $d$ is shown on a logarithmic scale). \note{Spectral loss is rescaled so that its value for one variable is equal to one.}
  (d) Eigenvalues $\pset{\lambda_{dk}}$ ($k$ is the index of the eigenvalue) calculated for the partial representations consisting of $d=1, 2, 3, 16$ variables. The eigenvalues of the Markov transition matrix from the complete set (magenta) are identical to the representation of $d=16$ variables (blue).
  (c) Eigenfunctions $\psi_1$ and $\psi_2$ colored as a function of the equilibrium eigenfunction $\psi_0$ (colorbar) shown for the partial selection $d=16$ and the complete data set $d=36$. \note{The eigenfunctions are calculated from the $d$-dimensional representation.}
  (d) Distributions of the $\psi_0$ and $\psi_1$ eigenfunctions obtained based on the partial ($d=16$) and complete selection ($d=36$) of the configuration variables.
  }
  \label{fig:cln}
\end{figure*}

We can also notice that for $d=10$, the eigenvalues match precisely the eigenvalues of the Markov transition matrix calculated on the complete data set $X$; see \rfig{fig:ala1}(d). Additionally, none of the selected configuration variables for $d=10$ describe the distances within the same residue, which means that the method picks only those distances relevant for describing the conformational transitions in alanine dipeptide. Instead, four correspond to the distances between Ace and Ala, five describe the distances between Ala and Nme, and one is between Ace and Nme. A list of the selected variables can be seen in Table S1 in Supplementary Information.

As a final example, we consider the folding and unfolding of a ten-residue protein chignolin in solvent; see inset in \rfig{fig:cln}(a). As before, we perform calculations using the \textsc{gromacs} 2019.2 code~\cite{gromacs} patched with a development version of the \textsc{plumed}~\cite{plumed,plumed-nest} plugin. We run a 1-$\mu$s well-tempered metadynamics~\cite{barducci2008well} simulation at 340 K with a bias factor of 5 and select samples every 8 ps from the last 20 ns of the simulation, amounting to the data set consisting of $N=2500$ samples. As biased variables to enhance transitions between the folded and unfolded states of chignolin, we choose the distance between C$\alpha$ atoms of residues Y1 and Y10 and the radius of gyration. The statistical weights are calculated using the Tiwary--Parrinello reweighting algorithm~\cite{tiwary_rewt}. From the resulting trajectory, we calculate the sines and cosines of the backbone $\Phi$ and $\Psi$ dihedral angles of chignolin and use them as the complete high-dimensional representation ($n=36$ variables in total).

We present the results in \rfig{fig:cln}. The spectral loss calculated for every partial selection of the configuration variables is shown in \rfig{fig:cln}(a). We can see that the spectral loss gradually decreases, reaching similarity with the complete selection for $d>15$. This fact can also be observed in \rfig{fig:cln}(b), where the eigenvalues of the Markov transition matrix calculated for $d=16$ are converged in reference to the eigenvalues computed from the complete data set. Moreover, the eigenfunctions $\psi_1$ and $\psi_2$ shown as a function of the equilibrium eigenfunction $\psi_0$ calculated based on the selections of the configuration variables for $d=16$ and $d=36$ are very similar; see \rfig{fig:cln}(c). Additionally, in \rfig{fig:cln}(d), we show that the distributions of the $\psi_0$ and $\psi_1$ eigenfunctions for the partial selection of the configuration variables ($d=16$) and the complete representation ($d=32$) are also in agreement. Our calculations show that less than half of the configuration variables in the complete high-dimensional representation is needed to represent the system. Interestingly, the configuration variables selected for $d=16$ correspond to every residue of chignolin, showing that the method identifies the variables that can carry the information about the folding and unfolding of chignolin while neglecting the possibly spurious variables. A list of the selected variables can be seen in Table S2 in Supplementary Information.

In this Letter, we present a simple and practical method for selecting high-dimensional representations of complex physical systems sampled by standard and enhanced sampling atomistic simulations. \note{The presented technique is general and requires only simulation data with statistical weights if the data is generated using an enhanced sampling technique. We use well-tempered metadynamics here for enhanced sampling and the Tiwary--Parrinello reweighting to generate data. However, data can be obtained from, for instance, parallel tempering, which does not require defining variables to bias before the simulation and is easy to reweight. In general, the presented algorithm can be tailored to any enhanced sampling technique; however, a reweighting scheme should be chosen as appropriate for the used method.}

The method constructs unbiased Markov transition matrices from simulation data and calculates their spectral decomposition for combinations of the configuration variables comprising the complete high-dimensional representation of the system. The selection algorithm creates a high-dimensional representation by iteratively removing the configuration variables from the data so that the spectral decomposition performed on the partial selection of the configuration variables resembles that of the complete representation. \note{This kinetic equivalence is obtained by minimizing the spectral loss that measures the deviation between the eigenvalues obtained from the considered high-dimensional representations. Since the selected partial high-dimensional representation is interpretable and preserves the timescale separation of the complete representation of the system, it can be used as an initial representation for subsequent construction of the slow variables, ensuring that the high-dimensional representation includes all relevant information about the system dynamics. When the timescale separation is ensured in a partial high-dimensional representation, further dimensionality reduction has the essential information about the dynamics of the studied complex physical system.} In conclusion, our method can become a helpful approach to analyzing high-dimensional physical systems and has the potential to be further explored.

\begin{suppinfo}
Supporting Information is available free of charge at \url{https://pubs.acs.org/}.
\begin{itemize}
  \item Details about simulations and data sets.
  \item \note{Details about constructing Markov transition matrices.}
  \item Selected configuration variables for high-dimensional systems.
\end{itemize}
\end{suppinfo}

\section*{Acknowledgements}
Funding from the Polish Science Foundation (START), the National Science Center in Poland (Sonata 2021/43/D/ST4/00920, ``Statistical Learning of Slow Collective Variables from Atomistic Simulations''), and the Ministry of Science and Higher Education in Poland is acknowledged. We thank Omar Valsson for critically reading the manuscript and Ming Chen for valuable discussions.

\bibliography{main}

\end{document}